\documentclass[9pt]{article}
\usepackage{amsfonts}
\usepackage{pstricks}
\usepackage{pst-node}
\usepackage{epsfig}
\usepackage{epsfig}
\usepackage[latin5]{inputenc}
\usepackage[T1]{fontenc}
\usepackage{lipsum}

\begin{document}
                \def\ba{\begin{eqnarray}}
                \def\ea{\end{eqnarray}}
                \def\w{\wedge}
                \def\d{\mbox{d}}
                \def\D{\mbox{D}}
 
%%%%%%%%%%%%%%%%%%%%%%%%%%%%%%%%%%%% TITLEPAGE %%%%%%%%%%%%%%%%%%%%%%%%%%%%%%
\begin{titlepage}            
\title{Non-Minimally Coupled Einstein-Yang-Mills Field Equations  and  Wu-Yang Monopoles in Bertotti-Robinson Spacetimes}
\author{Tekin Dereli\footnote{tdereli@ku.edu.tr},    
Yorgo \c{S}eniko\u{g}lu\footnote{ysenikoglu@ku.edu.tr}
\\
{\small  Department of Physics, Ko\c{c} University, 34450 Sar{\i}yer-\.{I}stanbul, Turkey}  }
%\affiliation{Department of Physics, Ko\c{c} University, 34450 Sar{\i}yer-\.{I}stanbul, Turkey }
\maketitle         
%\vskip 1cm
 
%\date{4 September 2019}
               
%vskip 2cm
 
\begin{abstract}
\noindent Bertotti-Robinson spacetimes are topologically $AdS_2 \times S^2$ and described by a conformally flat metric. 
Together with the Coulomb electric potential, they provide a class of static, geodetically complete Einstein-Maxwell solutions.
We show here that the Bertotti-Robinson metric together with Wu-Yang magnetic pole potentials give a class of static solutions of a 
system of non-minimally coupled  Einstein-Yang-Mills equations  that may be relevant for investigating vacuum polarization effects 
in a first order perturbative approach to quantum fields. 
 \end{abstract}
 
\vskip 1cm
 
 
\end{titlepage}
 
\newpage
%%%%%%%%%%%%%%%%%%%%%%%%%%%%%%%%%%%% INTRODUCTION %%%%%%%%%%%%%%%%%%%%%%%%%%%%%%
 
\section{Introduction}
Maxwell's  theory of electromagnetism is arguably one of the best established classical field theories of Nature.  
 Classically the coupling of electromagnetic fields to gravity are described by the Einstein-Maxwell field equations.
 This implies in the weak field approximation a quantum field theory of massless spin-1 photons coupled to massless spin-2 gravitons. 
 At the tree level the photon propagation takes place along the null geodesics of the Minkowski spacetime. 
 In perturbative QED in curved spacetimes though many unexpected new phenomena may occur. 
 At the 1-loop level there exists for instance vacuum polarization effects for which a photon produces a virtual electron-positron pair
that survives for a very short time interval before it pair annihilates back into a photon.
Such quantum vacuum  processes confer a length scale to the photon of the order of the Compton wavelength of an electron.
Thus a photon in motion in a curved spacetime will be affected by the gravitational fields, even at the 1-loop order. Perturbative effects such as these 
 in curved spacetimes can be conveniently described by an effective field theory at the classical level through certain non-minimal couplings of the electromagnetic fields to gravity\cite{B1,B2,B3,B4}. 

 In general, in an effective field theory defined at a mass scale $\mu$ , only those fields with masses lighter than or of the of order $\mu$ are taken into account.
 But the derivation of such an effective theory involves more than integrating out heavy particles of masses $M >\mu$ \cite{georgi,neubert}.
 Suppose one considers a renormalizable field theory at scales $\mu >> M$ with fields included at all the mass scales. 
 Then the theory is scaled down to  $\mu \approx M$ by the renormalization  group flow. 
 On the other hand, an effective field theory at scales $\mu <M$ with only those fields with masses less than or equal to $\mu$ is introduced 
 which may include higher derivative couplings and possibly many other types of non-renormalizable interactions. 
 Then the short-distance physics  that is incorporated into the coefficients of the effective Lagrangian 
is disentangled from the long-distance physics that remains explicit in the actual low energy theory. Such a disentanglement can be achieved order by order
 in perturbation theory by matching the physical parameters of the high energy theory at $\mu >>M$ to the corresponding parameters of the low energy theory at 
mass scales $\mu \approx M$.

In QED, it is found that when the electron field is integrated out, the resulting effective Lagrangian 
 at the 1-loop level involves certain  $F^4$-type self couplings of the electromagnetic fields. The corresponding long-distance theory complements Maxwell electrodynamics by the higher derivative  Heisenberg-Euler terms \cite{B5}:
 $$
 {\mathcal{L}}= \frac{1}{2} F \wedge *F + \frac{2\alpha^2}{45 m^4}  \left (X^2 + \frac{7}{4}Y^2 \right ) *1.
$$
In a similar way, the effective action density 4-form at the 1-loop level in a curved spacetime for QED had been calculated \cite{B7}:
\begin{eqnarray}
{\mathcal{L}}_{eff} &=&  \frac{12 \alpha}{45\pi m^2}  (d*F) \wedge *(d*F) + \frac{\alpha}{45\pi m^2} R_{ab} F^{ab} \w *F  \nonumber \\ & &-\frac{13\alpha}{45\pi m^2} Ric_{a} \wedge \iota^aF \wedge *F +\frac{5\alpha}{90\pi m^2} R F \wedge *F.  \nonumber
\end{eqnarray}
 where $\alpha$ is the fine structure constant and $m$ is the free electron mass.  The 2-photon 2-graviton vertices are settled in perturbative QED by the conventional $RF^2$ interactions in the effective action. These interaction vertices were found to mean photon graviton oscillations at low energies relative to the Planck scale and in weak gravitational fields \cite{B8}. This process may be significant in the presence of strong magnetic fields near heavy magnetars and pulsars \cite{B9,D1,D2}.
  In fact a curved spacetime acts as an optically active medium for the photon with an index of refraction determined by the higher-loop QED corrections \cite{shore3,dereli,hollowood}.  
Then dispersion effects such as gravitational rainbows or gravitational bi-refringence  may occur.

The distinctive feature of a non-Abelian gauge field theory is the self - interactions among its gauge bosons. In QCD, the exchange potentials between color charges give rise to asymptotically free forces. Then in the UV limit constituent quarks behave as if they are free particles inside the hadrons. Perturbative quantization is viable in this case and the high-energy QCD tests agree with observations up to logarithmic corrections. On the other hand the perturbation theory fails altogether in the IR limit where the interaction potential becomes very complicated. The corresponding low-energy limit of QCD is not easy to define. A possible approach to deal with such long distance effects of QCD is to work with an effective field theory that would be valid at lower energy scales. A perturbative 1-loop effective field theory for QCD includes the generic types of $F^3$ and $(\nabla_{A}F)^2$ Yang-Mills self-interaction terms that show up as exotic strong-interaction contributions in colour condensate models \cite{baskal-dereli}. Then it  should be expected in curved spacetimes, the QCD vacuum polarization effects shall also contribute, as it is in the case of perturbative QED, non-minimal coupling terms of the generic type $RF^2$ at the 1-loop level to an effective Einstein-Yang-Mills theory \cite{dereli-ucoluk}.  
 
 \medskip
 
 We consider in Section:2  a simple non-minimally coupled Einstein-$SU(2)$ Yang-Mills effective field theory model.
Variational field equations will be derived in the coordinate-independent language of exterior differential forms. 
We refer to our previous papers for the notation and conventions \cite{B10b,B7-e,B7-f}.
We provide in Section:3  a brief discussion of the unique conformally flat, static, spherically symmetric Bertotti-Robinson solution of the source-free Einstein-Maxwell equations. 
The celebrated Wu-Yang magnetic pole potentials are introduced in Section:4. 
In particular we give a non-generic class of  regular static, spherically symmetric solutions of the coupled Einstein-Yang-Mills field equations.

%%%%%%%%%%%%%%%%%%%%%%%%%%%%%%%%%%%%%%%%%%%%%%%%%%%%%%%%%%%%%%%%%%%%%%%%%%%%%%
 
\section{Non-minimally Coupled Einstein-Yang-Mills \\Fields}

\noindent We first consider the Yang-Mills potential 1-form\footnote{We set the gauge coupling constant $e=1$}
\begin{equation}
A=A^{j}T_{j} = A^{j}_{\;\;\mu} dx^{\mu} \otimes T_{j} 
\end{equation}
where $\{T_j  :  j=1,2,3\}$ are the anti-hermitian generators of the Lie algebra $su(2)$ that satisfy the structure equations
\begin{equation}
\left [ T_j , T_k \right ] = \epsilon_{jkl} T_{l}.
\end{equation}
One may also introduce a local coordinate chart $\{x^\mu : \mu=0,1,2,3 \}$, even though we will be using extensively the coordinate independent 
formalism of exterior differential forms over the spacetime. 
Then the Yang-Mills field 2-form is determined by 
\begin{equation}
F=dA + A\wedge A = F^{j} T_j = \frac{1}{2} F^{j}_{\;\;\mu \nu} dx^\mu \wedge dx^\nu \otimes T_{j}.
\end{equation} 
We work out 
$$
F^{j} = dA^j + \epsilon_{jkl} A^k \wedge A^l \;\; , \;\; 
 F^{j}_{\;\;\mu \nu} = \partial_{\mu} A^{j}_{\;\;\nu} -  \partial_{\nu} A^{j}_{\;\;\mu} + \epsilon_{jkl} A^{k}_{\;\;\mu}A^{l}_{\;\;\nu}.
 $$
Under a local gauge transformation $U = e^{\theta^{j}T_{j}}  \in SU(2)$,  the Yang-Mills fields transform as
$$
A \rightarrow U A U^{-1} + U dU^{=1} \quad , \quad F \rightarrow U F U^{-1} 
$$
The integrability of $F$  implies the Bianchi identity
\begin{equation}
\nabla_{A}F \equiv dF + A \wedge F - F \wedge A =0.
\end{equation}
We label the quadratic Yang-Mills invariants by
\begin{equation}
X = *\hspace{0.5mm}Tr(F\wedge *F), \quad Y = *\hspace{0.5mm}Tr(F\wedge F).
\end{equation}

\noindent The Lagrangian density 4-form of Einstein-Yang-Mills theory with a cosmological constant is given by
\begin{equation}
{\mathcal{L}}_0 = \left ( \frac{1}{2\kappa^2} R   + \frac{1}{2} X  + \Lambda \right ) *1
\end{equation}
where 
\begin{equation}
R =-*(R_{ab} \wedge *e^{ab}).
\end{equation}
denotes the scalar curvature of space-time. $\kappa^2=4\pi G$ with $G$ being the Newton's universal constant and $\Lambda$ is a cosmological constant.

\noindent We consider, for the sake of simplicity,  only the following non-minimal coupling terms:
\begin{equation}
{\mathcal{L}}_1 = \frac{\gamma}{2} X R *1  +  \frac{\gamma^\prime}{2} Y R *1.
\end{equation}
$\gamma$ and $\gamma^{\prime}$ are two arbitrary coupling constants.
Furthermore we assume that the connection is the unique, torsion-free Levi-Civita connection. 
This constraint will be imposed by   the method of Lagrange multipliers
\begin{equation}
{\mathcal{L}}_C = \left (de^a + \omega^{a}_{\;\;b} \wedge e^b \right ) \wedge \lambda_a.
\end{equation}
Therefore the total action that is to be  varied becomes
\begin{equation}
I[ e^a, \omega^{a}_{\;\;b}, A, \lambda_a] = \int_{M} \left (  {\mathcal{L}}_0 + {\mathcal{L}}_1 + {\mathcal{L}}_C  \right )
\end{equation}
After a long computation, the final form of the  gravitational field equations turn out to be
\begin{eqnarray}
( \frac{1}{\kappa^2} + \gamma X  + \gamma^{\prime} Y  ) G_a &-& (\Lambda -\frac{\gamma}{2} X R -\frac{\gamma^\prime}{2} Y R  ) *e_a  \\
&=& \left (  1+\gamma R \right ) \tau_{a}[F] + \gamma D( \iota_a *dX)  + \gamma^{\prime}  D( \iota_a *dY),  \nonumber  \label{einstein}
\end{eqnarray}
where the Einstein 3-forms
\begin{equation}
G_a = -\frac{1}{2} R^{bc} \wedge *e_{abc} = G_{ab} *e^b  \nonumber
\end{equation}
and the Yang-Mills stress-energy-momentum 3-forms
\begin{equation}
\tau_{a}[F] = \frac{1}{2}Tr \left ( \iota_a F \wedge *F - F \wedge \iota_a *F \right ) = T_{ab}[F] *e^b .
\end{equation}
The Yang-Mills field equations are also modified according to
\begin{equation}
\nabla_{A}F=0, \quad  \nabla_{A}*\left ( F +\gamma R F + \gamma^\prime R *F \right ) =0.   \label{maxwell}
\end{equation}
 In what follows we will be looking for the solutions of the coupled field equations (11) and (14).
 
%%%%%%%%%%%%%%%%%%%%%%%
 
\section{Bertotti-Robinson Spacetimes}

\noindent The Bertotti-Robinson spacetimes \cite{bertotti,robinson,stephani,dolan,tariq-tupper,dadhich,garfinkle-glass} admit a product topology
$ M^{1,3} = AdS_2 \times S^2.$ 
They were first discussed as the unique conformally flat, static spherically symmetric solution 
of the source-free Einstein-Maxwell equations with a non-null electromagnetic potential.
In particular, we consider  in spherical polar coordinates $(t,r,\theta, \varphi)$  a static, spherically symmetric  metric 
\begin{equation}
g = -f_{1}^2(r) dt^2 + f_{2}^2(r) dr^2 + r_{0}^2 \left ( d\theta^2 + \sin^2 \theta d\varphi^2 \right ),
\end{equation}
together with  a static, spherically symmetric electric potential
\begin{equation}
A = q(r) dt.
\end{equation}
Then the source-free Einstein-Maxwell field equations 
\begin{equation}
-\frac{1}{2} R^{bc} \wedge *e_{abc}  = \frac{\kappa^2}{2} \left ( \iota_aF \wedge *F - F \wedge \iota_a*F \right ),
\end{equation}
\begin{equation}
dF=0 \quad, \quad d*F=0
\end{equation}
are satisfied for the choice 
$$
f_{1}(r) = f_{2}(r) =\frac{r_0}{r} \quad, \quad q(r) = -\frac{Q}{r} .
$$    
Hence the metric   
\begin{equation}
g = \frac{r_0^2}{r^2} \left [  -dt^2 + dr^2 + r^2 (d\theta^2 +\sin^2\theta d\varphi^2) \right ]
\end{equation}
is conformally flat and the Weyl curvature 2-forms identically vanish:
\begin{equation}
C_{ab} = R_{ab} -\frac{1}{2} ( e_a \wedge Ric_b - e_b \wedge Ric_a ) +\frac{R}{6} e_a \wedge e_b =0.
\end{equation}
The curvature scalar $R=0$ vanishes as well. $A$ turns out to be the Coulomb electric potential so that the net electric charge 
\begin{equation}
\frac{1}{4\pi} \oint_{S^2} *F = Q.
\end{equation}

\noindent We also make note of an alternative way of writing down the Bertotti-Robinson solution with a re-definition of the time coordinate as
\begin{equation}
dt =(\frac{r}{r_0})^2 d\tilde{t}. 
\end{equation}
Then the conformally flat metric reads
\begin{equation}
g = -\left (\frac{r^2}{r_0^2} \right )^2 d{\tilde{t}}^2 + \left (\frac{r_0^2}{r^2} \right )^2 \left [ dr^2 + r^2 (d\theta^2 + \sin^2\theta d\varphi^2 ) \right ],
\end{equation}
and the electric potential 1-form becomes
\begin{equation}
A= -\frac{Q}{r_0^2} r d{\tilde{t}} .
\end{equation}

\section{Wu-Yang Magnetic Pole Solutions}
We now look for static, spherically symmetric solutions of our Einstein-Yang-Mills system of equations for which the space-time metric  in terms of isotropic coordinates $\{t,x,y,z\}$ 
is given by  
\begin{equation}
g = - \left ( \frac{r}{r_0}\right )^{2k} dt^2 + \left ( \frac{r_0}{r}\right )^{2} \left ( dx^2 + dy^2 + dz^2 \right ),
\end{equation}
where  $r=\sqrt{x^2+y^2+z^2}$.
We introduced a real parameter  $k$ so that our Bertotti-Robinson metric  is no longer conformally flat unless $k^2=1$. 
Yet  the spacetime still admits a product topology $AdS_2 \times S^2$.
We calculate  the curvature scalar which turns out to be 
\begin{equation}
R =-\frac{2}{r_0^2}(1-k^2). 
\end{equation}

\noindent Let us take the Wu-Yang potential 1-forms \cite{wu-yang,bartnik-mckinnon,balakin1}
\begin{eqnarray}
A^1 = \left ( \frac{K(r) - 1}{r^2} \right ) (ydz-zdy), \nonumber \\
A^2 =  \left ( \frac{K(r) - 1}{r^2} \right )(zdx-xdz), \nonumber \\
A^3 =  \left ( \frac{K(r) - 1}{r^2} \right ) (xdy-ydx).
\end{eqnarray}
Then we calculate
\begin{eqnarray}
F^1&=&\big[(y^2+z^2) \frac{1}{r} \left ( \frac{K(r) - 1}{r^2} \right )^{\prime} +2  \left ( \frac{K(r) - 1}{r^2} \right ) + 2e^2 x^2   \left ( \frac{K(r) - 1}{r^2} \right )^2\big]dy \wedge dz \nonumber \\
&&+\left [ \frac{1}{r}\left ( \frac{K(r) - 1}{r^2} \right )^{\prime} -2 \left ( \frac{K(r) - 1}{r^2} \right )^2 \right ] xdx \wedge (ydz - zdy), 
\end{eqnarray}
and by the cyclic permutations of $(xyz)$, we also write down 
\begin{eqnarray}
F^2&=&\big[(z^2+x^2) \frac{1}{r} \left ( \frac{K(r) - 1}{r^2} \right )^{\prime} +2  \left ( \frac{K(r) - 1}{r^2} \right ) + 2e^2 y^2   \left ( \frac{K(r) - 1}{r^2} \right )^2\big]dz \wedge dx \nonumber \\
&&+\left [ \frac{1}{r}\left ( \frac{K(r) - 1}{r^2} \right )^{\prime} -2 \left ( \frac{K(r) - 1}{r^2} \right )^2 \right ] ydy \wedge (zdx - xdz), 
\end{eqnarray}
and 
\begin{eqnarray}
F^3&=&\big[(x^2+y^2) \frac{1}{r} \left ( \frac{K(r) - 1}{r^2} \right )^{\prime} +2  \left ( \frac{K(r) - 1}{r^2} \right ) + 2e^2 z^2   \left ( \frac{K(r) - 1}{r^2} \right )^2\big]dx \wedge dy \nonumber \\
&&+\left [ \frac{1}{r}\left ( \frac{K(r) - 1}{r^2} \right )^{\prime} -2 \left ( \frac{K(r) - 1}{r^2} \right )^2 \right ] zdz \wedge (xdy - ydx). 
\end{eqnarray}
Suppose special solutions of the Yang-Mills equations for which $K(r)=K$ is a constant are sought. In this case we calculate
\begin{equation}
d*F^1 + A^2 \wedge *F^3 - A^3 \wedge *F^2 = \left (  \frac{r}{r_0} \right )^{k+1} \frac{ K(K^2 - 1)}{ r^3} dt \wedge dr  \wedge dx =0
\end{equation}
and the other two field equations are similarly determined by cyclic permutations of $(123)$ and $(xyz)$. 
There exists three constant solutions:  $K=\pm1$ and $K=0$. In the first two cases the Yang-Mills fields vanish, $F=0$, so that  both of these cases correspond to the Yang-Mills vacuum.  The non-trivial Wu-Yang magnetic pole solution corresponds to the choice $K=0$.
On the other hand, substituting in the Yang-Mills field 2-forms above and tracing over the Lie algebra, we get 
\begin{equation}
X =   -\left ( \frac{r}{r_0} \right )^4 \Big( \frac{{2K'}^2}{r^2}+\frac{ (K+1)^2(K-1)^2}{r^4} \Big), \quad Y =0.
\end{equation}
 Therefore it can easily be verified for the Wu-Yang solution that  
\begin{equation}
X=-\frac{1}{r_0^4}, \quad R=-\frac{2}{r_0^2}(1-k^2).
\end{equation}
Finally the complete set of non-minimally coupled  Einstein-Yang-Mills field equations with a cosmological constant are solved provided
\begin{equation}
\Lambda = \frac{1}{2r_0^4} \quad , \quad \gamma = \frac{r_0^4}{\kappa^2} .
\end{equation}
Then the metric parameter should be bound according to 
\begin{equation}
0 < k^2 = 1 - \frac{\kappa^2}{2 r_0^2} < 1.
\end{equation}
It is interesting to note that under the assumption that the coupling constants are fixed by
\begin{equation}
\kappa^2 \Lambda =\frac{1}{2 \gamma } =\frac{1}{2 \gamma^{\prime} }, 
\end{equation}
the total Lagrangian density 4-form factorizes as follows:
\begin{equation}
{\mathcal{L}}_0 + {\mathcal{L}}_1 = \left ( \frac{1}{2\kappa^2} R + \Lambda \right ) \left ( 1+ \gamma \kappa^2 X + \gamma^{\prime}\kappa^2 Y \right ) *1.
\end{equation}
For the solution above, both of these factors vanish. Hence we get a very special type of an exact solution of the Einstein-Yang-Mills field equations
derived from a Lagrangian density that admits a product structure. Then
under independent field variations with respect to the metric and the gauge potential, the variational field equations will always contain either the first factor or the second factor. 
For our solution, it is  evident that the coupled field equations are satisfied  since  both the gravity and gauge factors vanish on their own.

\section{Concluding Remark}
Here a simple model of non-minimally coupled Einstein-Yang-Mills field equations are considered. A very special class of exact static, spherically symmetric 
solutions which are determined by the Wu-Yang magnetic pole potentials in a non-conformally flat Bertotti-Robinson spacetime are found. 
Such solutions may be related physically to regular black holes
with colour \cite{balakin2,liu et al, zhang-mann}. However, they are  in no way generic. We regard the existence  of these non-generic solutions in a simple 
case of  non-minimal couplings induced by 1-loop quantum effects in curved spacetimes as an indication that 
 Yang-Mills vacuum polarization effects might in general imply unexpected new results. 
 
\section{Acknowledgement}
Y.\c{S}. is grateful to Ko\c{c} University for its hospitality and partial support.

\end{document}